\documentclass[12pt]{article}

\usepackage{scicite}
\usepackage[labelfont=bf,labelsep=space]{caption}
\usepackage{times}
\usepackage{cite}
\usepackage{color,soul}
\usepackage{graphicx}

\newenvironment{sciabstract}{%
\begin{quote} \bf}
{\end{quote}}

\title{Coherent Transfer of Lattice Entropy via Extreme Nonlinear Phononics in Metal Halide Perovskites}

\author
{Z. Liu$^{1}$, Y. Shi$^{2}$, T. Jiang$^{1}$, L. Luo$^{1}$, C. Huang$^{1}$, M. Mootz$^{1}$, Z. Song$^{3}$, Y. Yan$^{3}$, \\
	Y. Yao$^{1}$, J. Zhao$^{2}$,
	J. Wang$^{1\dag}$
	\\
	\normalsize{$^{1}$Ames National Laboratory, Ames, IA 50011 USA,}\\
\normalsize{and Department of Physics and Astronomy, Iowa State University,}\\
\normalsize{Ames, IA 50011, USA}\\
\normalsize{$^{2}$ICQD/Hefei National Laboratory for Physical Sciences at the Microscale,}\\
\normalsize{and Key Laboratory of Strongly-Coupled Quantum Matter Physics,}\\
\normalsize{Chinese Academy of Sciences,}\\
\normalsize{and Synergetic Innovation Center of Quantum Information $\&$ Quantum Physics,}\\
\normalsize{University of Science and Technology of China, Hefei, Anhui 230026, China}\\
\normalsize{$^{3}$Department of Physics and Astronomy and Wright Center for Photovoltaics} \\
\normalsize{Innovation and Commercialization, The University of Toledo, Toledo, OH 43606, USA.}\\
\normalsize{$^\dag$To whom correspondence should be addressed; E-mail: jgwang@ameslab.gov.}
}
\date{}

\begin{document} 


\baselineskip24pt


\maketitle 
\begin{sciabstract}
Entropy transfer in metal halide perovskites, characterized by significant lattice anharmonicity and low stiffness, underlies the remarkable properties observed in their optoelectronic applications, ranging from solar cells to lasers. The conventional view of this transfer involves stochastic processes occurring within a thermal bath of phonons, where lattice arrangement and energy flow from higher to lower frequency modes.
Here we unveil a comprehensive chronological sequence detailing a conceptually distinct, coherent transfer of entropy in a prototypical perovskite CH$_3$NH$_3$Pbl$_3$. 
The terahertz periodic modulation imposes vibrational coherence into electronic states, leading to the emergence of mixed (vibronic) quantum beat between approximately 3~THz and 0.3~THz.
We highlight a well-structured, bi-directional time-frequency transfer of these diverse phonon modes, each developing at different times and transitioning from high to low frequencies from 3 to 0.3~THz, before reversing direction and ascending to around 0.8~THz.
First-principles molecular dynamics simulations disentangle a complex web of coherent phononic coupling pathways and identify the salient roles of the initial modes in shaping entropy evolution at later stages. 
Capitalizing on coherent entropy transfer and dynamic anharmonicity presents a compelling opportunity to exceed the fundamental thermodynamic (Shockley-Queisser) limit of photoconversion efficiency and to pioneer novel optoelectronic functionalities.
\end{sciabstract}

$\newline$
Recently, there has been a growing relevance of quantum coherence and non-equilibrium dynamics in optoelectronic materials exploration and energy conversion systems \cite{n1,n2, newadd2,newadd3,newadd4,newadd7,c1,c2}. These phenomena can have immediate and profound effects on entropy transport and energy conversion efficiencies. 
In contrast to the conventional, incoherent diffusive transfer processes, the comprehension of coherent and reversible transfer of structural and electronic excitations remains limited. 
This research enables the deliberate design of initial pumping processes, which can then be harnessed to guide and manipulate the subsequent evolution of energy and entropy in a controlled manner. Particularly, entropy transfer reflects the distribution of energy among the vibrational modes of the lattice, placing emphasis on the diverse atomic arrangements within the lattice structure. In this regard, the remarkable lattice anharmonicity \cite{liu2, newadd5,newadd9} from highly polarizable organic cations and inorganic lattice found in hybrid organic-inorganic metal halide perovskites, such as methylammonium lead iodide (MAPbI$_3$), make them an ideal model system for investigating and comprehending the novel concept of coherent entropy transfer. 

Hybrid organic-inorganic perovskites have garnered significant attention in recent years owing to their remarkable properties, positioning them as one of the most intriguing optoelectronic materials \cite{ref5,ref7, newadd10,newadd11,newadd12, liu1, liu2, liu3, luo, di} for potential device applications \cite{snaith}.
Remarkable performance across various applications spans from light-emitting diodes (LEDs) and lasers to highly efficient solar cells \cite{xiao}. These devices leverage the distinctive structure of metal halide perovskites, with MAPbI$_3$ serving as a prime example. This structural motif embodies a resilient inorganic octahedral framework, PbI$_6$, synergistically coupled with a lightweight organic group, CH$_3$NH$_3$. In contrast to inorganic semiconductors, perovskites exhibit a lattice structure characterized by low stiffness and pronounced anharmonicity\cite{liu2, ref5}, 
which plays a considerable role in the fundamental optoelectronic and charge transport properties. 
The lattice structure is found to host numerous THz phonon modes, which have been identified through techniques such as Raman spectroscopy\cite{Frost} and infrared absorption measurements\cite{ref6}. However, the time sequence of the coherent transfer of entropy among these phonon modes remains a largely uncharted territory, which, therein, presents a major next challenge.

Here we discover bi-directional, coherent entropy transfer through THz-driven vibronic, or vibrational-electronic coupled, quantum beat spectroscopy in MAPbI$_3$. This is accomplished by detecting excitonic level oscillations induced by the periodic modulation of specific lattice modes through a non-contact THz ``push" electric field with minimum heating. 
By employing a transformation into the time-frequency domain of the THz-driven quantum beats, we directly observe and track the various lattice vibrations in their chronological sequence. Our discovery highlights two unique features: a bi-directional coherent transfer mechanism, facilitating energy flow from higher to lower frequency phonons, then reverses back to higher frequencies, contrasting with the usual incoherent and unidirectional transfer; an exceptionally strong nonlinear phononic coherent coupling spanning from frequencies $\sim$3 THz down to 0.3 THz, $\sim$15 times larger than the phonon shifts typically observed in other representative semiconductors such as Ti$_2$O$_3$.  
These results, together with first-principles molecular dynamics (FPMD) simulations, demonstrate the entropy transfer pathway sensitive to the initial excitation of phonon modes. 
\begin{figure*}[tbp]
	
	\includegraphics[scale=0.6]{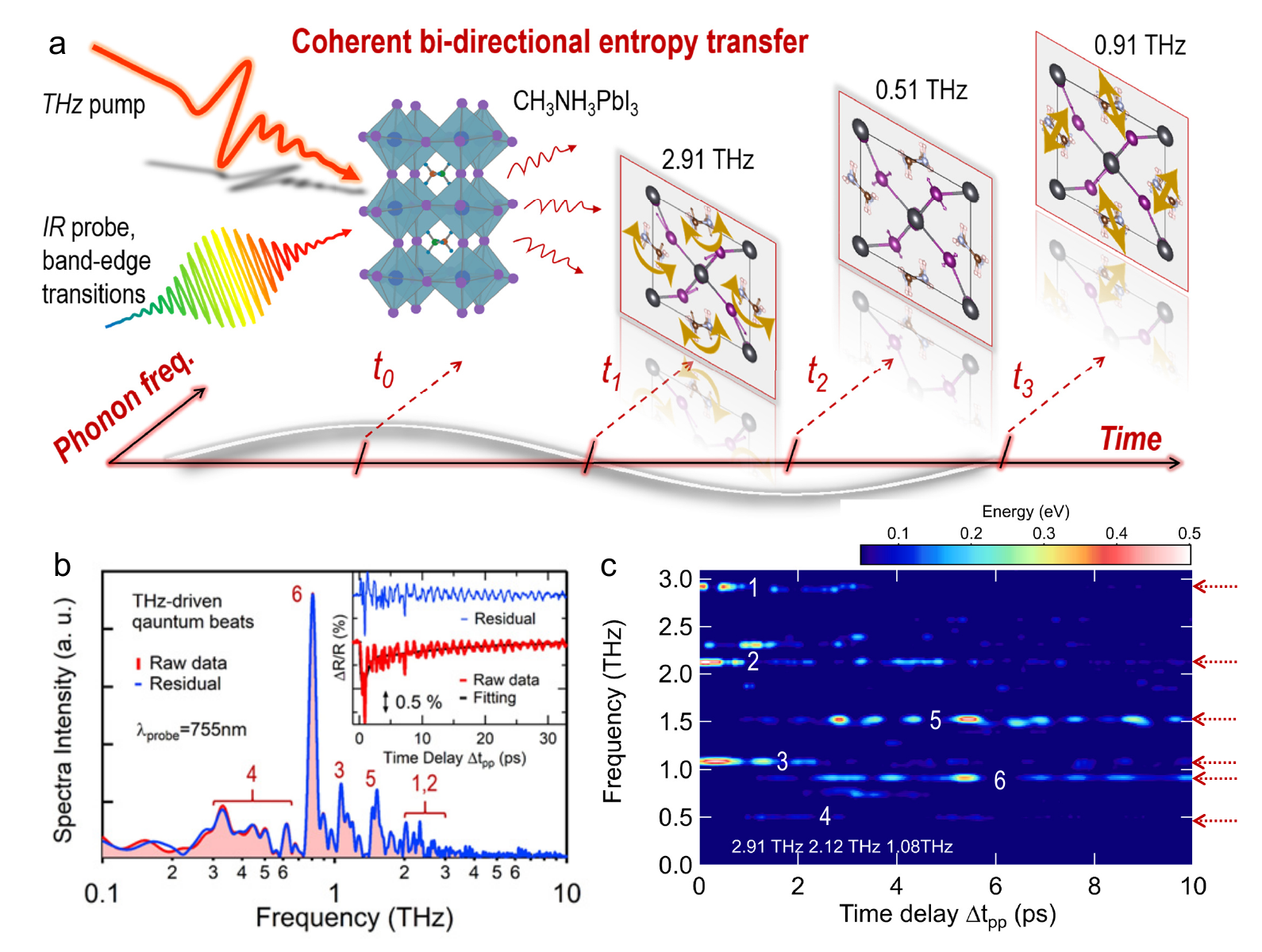}
	\caption{(a) An illustration of a nonlinear entropy transfer pathway driven by the strong THz pulse radiation. The pathway is delineated through three representative phonon modes as computed below, sequentially spanning from high to low frequency (2.91 THz to 0.51 THz), then reverting back to high frequency (0.91 THz). (b) Fourier spectra of the transient reflectivity $\Delta R/R$ probe of vibronic, or mixed (excitonic-vibrational), oscillation (red shade) and its oscillatory residual $\Delta R_{osc}/R$  (blue line) at 755 nm probe wavelength at 4.2 K. Phonon modes most relevant are labeled by the numbers 1-6. Inset: $\Delta R/R$ dynamics of MAPbI$_3$ at 755 nm (red) at 4.2 K. Shown together is the fitted exponential component (black, Supplemental Note 3) and the corresponding oscillatory residual $\Delta R_{osc}/R$ (blue). (c) The time-evolution of the system's phonon mode-selective energy distribution function over the 10 ps FPMD window is plotted, with three initially populated phonons ($\omega_1$, $\omega_2$, and $\omega_3$) at frequencies of 2.91, 2.12, and 1.08 THz, respectively, guided by the experimental findings depicted in Fig. 2. The color scale denotes the energy associated with each phonon mode.}
	\label{Fig1}
\end{figure*}

Our experimental schematics of intense THz pump and white light probe spectroscopy is illustrated in Fig. 1a (Supplemental Note 1). 
A strong single-cycle THz pump pulse centered at 1.5 THz (6.2 meV) with peak E-field $E_{THz}\sim$1000
~kV/cm \cite{s1,s2,s3} was used to excite the MAPbI$_3$ single crystal (Supplemental Note 2) in its low-temperature orthorhombic phase.
The use of intense THz pulses has proven to be a powerful tool for exciting THz coherent phonons of both infrared (IR) and Raman symmetries in complex quantum materials \cite{m1,m2,1}, including metal halide perovskites \cite{liu1, newadd1,newadd8}. 
In our experiments with the MAPbI$_3$ sample, the intense THz pumping serves as the initial trigger for a nonlinear coherent entropy transfer process. We subsequently employed a white light probe pulse with a duration of approximately $\sim100$ femtoseconds to detect the excitonic transition. The probe pulse was adjustable across a wavelength spectrum ranging from 730 nm (1.7 eV) to 773 nm (1.6 eV), effectively encompassing the excitonic bands present in MAPbI$_3$ \cite{liu1,liu2}. 
Our THz pump and excitonic transition probe spectroscopy approach enables the measurement of mixed (vibronic) quantum beats arising from coherent phonon-modulated electronic states.

The pump-induced coherent oscillations at the exciton transitions were uncovered by examining the transient differential reflectivity $\Delta R/R$ trace. The inset of Fig. 1b shows a representative $\Delta R/R$ trace (red solid trace) at the probe wavelength of 755 nm at 4.2 K as a function of time delay $\Delta t_{pp}$ up to $\sim$ 33 ps. 
We notice a long-lasing oscillatory component $\Delta R_{osc}/R$ of more than 10s of ps (solid blue line in the inset) is superimposed on the exponential one (solid black line in the inset), 
which can be very well fitted (see Supplemental Note 3). 
The exponential decay characterizes lifetime of the excitons induced by THz pump-induced tunneling ionization, while the oscillatory signals elucidate the presence of vibronic coherence, indicating mixed exciton-phonon coherence.
In order to reveal the intrinsic quantum beat spectra \cite{q1}, we extract the oscillatory residual component $\Delta R_{osc}/R$ (blue solid line) by removing the exponential decay component.
The corresponding Fourier transform (FT) spectra of $\Delta R_{osc}/R$ dynamics (blue) and $\Delta R/R$ (red) are shown in Fig.1b, respectively. 
The FT spectrum of $\Delta R_{osc}/R$ provides a wealth of phonon modes, as identified and marked with red labels as $\omega_1$ through $\omega_6$ in Fig 1b, with several notable features. First, a prominent transverse optical phonon mode denoted as $\omega_6$ due to the twist of octahedra cage is clearly observed at 0.8 THz\cite{Frost}. Two strong secondary peaks are evident at approximately 1.08 THz ($\omega_3$) and 1.52 THz ($\omega_5$), respectively. Second, multiple peaks are distributed within a phonon band ranging from 0.3 to 0.7 THz, as indicated by the label $\omega_4$, which are shown to be related to the MA molecular rotations\cite{ref1,moving}. Third, finally, a couple of distinct high-frequency phonon modes, marked as $\omega_1$ and $\omega_2$, appear in the range of 2 to 3 THz. 
The FT spectrum of coherent dynamics reveals an extraordinarily nonlinear phononics and represents a starting point for investigating the coherent transfer of various structural vibrations, each of which develops at significantly different times. 

Next, our focus shifts to the measurement of highly nonlinear phononic coupling processes, distinguished by coherent and bi-directional entropy transfer. 
	As depicted in Fig. 1a, a prominent phonon transfer process is discerned, involving three representative modes transitioning from 2.91 THz to 0.51 THz and then back to 0.91 THz, evolving sequentially. This observation is substantiated by both theoretical predictions and experimental results below.
To gain insights into the temporal development and evolution of the diverse phonon modes depicted in Fig. 1b, we initiated our investigation by examining the predictions from the FPMD simulations shown in Fig. 1c (Supplemental Note 5). We strategically selected specific phonon modes to serve as initial conditions, i.e., phonon modes $\omega_1$ at 2.91 THz and $\omega_2$ at 2.12 THz to represent the high energy phonons, and a mode $\omega_3$ at 1.08 THz observed in Fig. 1b. The selection of these specific phonon modes was guided by experimental observations presented in Fig. 2.

We first present the simulation results in Fig. 1c, offering a clear and compelling illustration of the gradual buildup and coherent evolution of diverse phonon modes within MAPbI$_3$, as indicated by red dashed arrows and white labels.
The first ps of the simulation shows that  the intensities of modes $\omega_1$ and $\omega_2$ decrease, whereas mode $\omega_3$ experiences enhancement, partially attributed to the transfer of energy from modes $\omega_1$ and $\omega_2$. This behavior aligns with expectations based on the ionic difference-frequency-type excitation mechanism \cite{ref4}. 
Following the initial picosecond, mode $\omega_3$ begins to diminish, with a portion of its energy transferred to facilitate the emergence of mode $\omega_4$ at 0.51 THz.
In the later stages, mode $\omega_5$ at 1.52 THz is induced, coinciding with the gradual weakening of mode $\omega_4$. This phenomenon stems from a sum-frequency-type generation mechanism. Eventually, a new mode, $\omega_6$, emerges at 0.91 THz and persists for the duration of the simulation.
Note that all the numerical results are obtained from Frist-principles calculations based on rigorous density functional theory (DFT) results, and some discrepancies between theoretical (Fig. 1c) and experimental (Fig. 1b) values of phonon frequencies, are commonly accepted. For instance, we designate the theoretical model with a frequency of 0.51 THz to correspond to the experimental $\omega_4$ mode depicted in Fig. 1b. It is noteworthy that the $\omega_4$ mode appears most prominently in Fig. 2 at 0.3 THz.
Remarkably, these findings surpass the conventional incoherent energy transfer, which typically involves a unidirectional flow from high to low energy phonons. These intricate dynamics underscore the coherence and interplay of various organic molecules and inorganic cage modes to drive the entropy transfer.

\begin{figure*}[tbp]

	\includegraphics[scale=0.55]{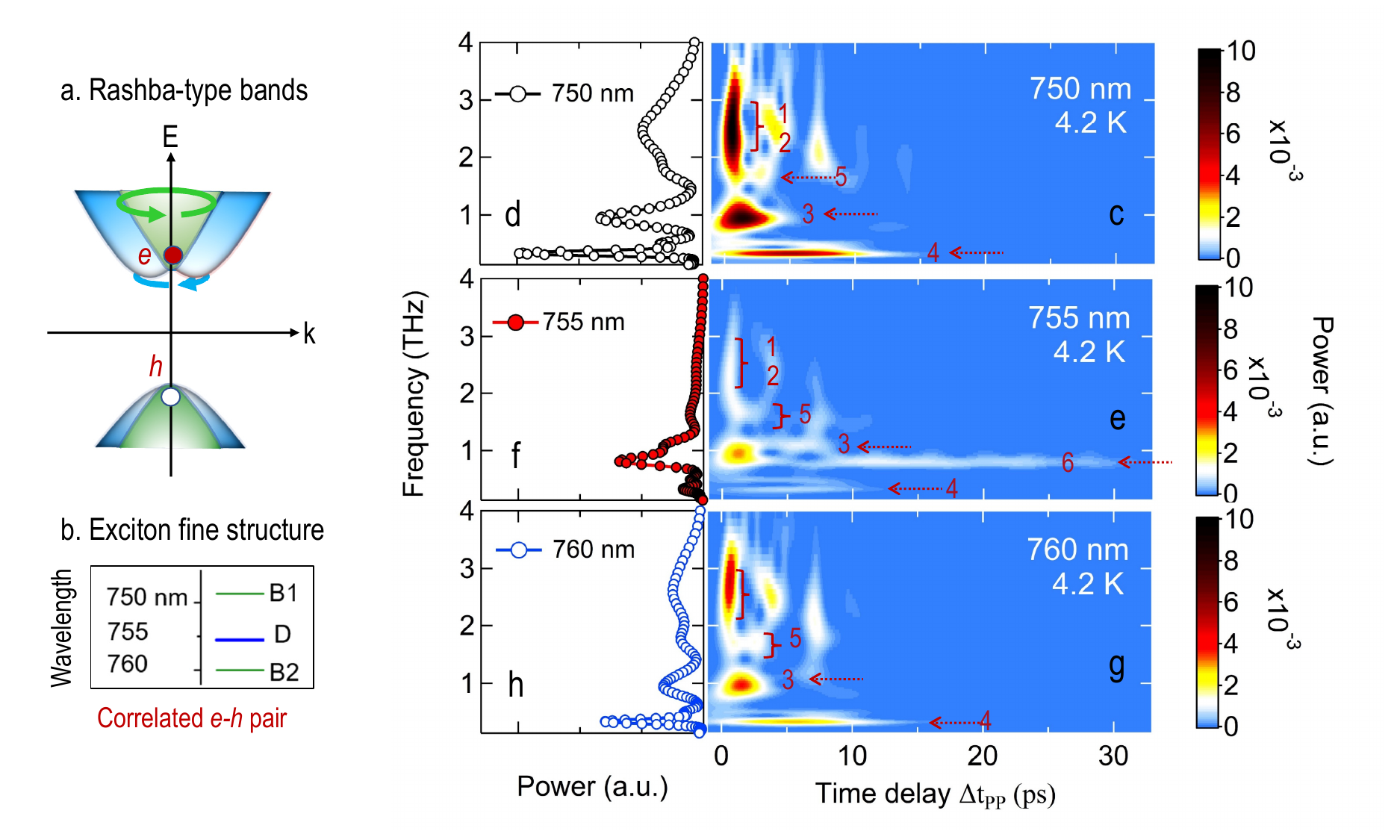}
	\caption{(a) An illustration of Rashba-type spin split bands in MAPbI$_3$. (b) Correlated electron-hole pairs within Rashba-type bands, showcasing excitonic fine structure where dark exciton states (labeled as D) reside between two brightly emitting exciton states (labeled as B1 and B2) within the spectral range of 750--760 nm.	Continuous wavelet transform of THz pump-driven
		$\Delta R_{osc}/R$ dynamics at probe wavelength 750 nm (c), 755 nm
		(e) and 760 nm (g) at 4.2 K. Shown together are the time-integrated
		power spectra from time-frequency mapping of the measured 750 nm (b,
		black), 755 nm (d, red ) and 760 nm (f, blue) traces. The representative phonon modes $\omega_1$--$\omega_6$ identified in Fig. 1b are labeled by the arrows and the numbers 1-6.}
	
\end{figure*}

To validate the highly nonlinear entropy transfer pathway predicted by the FPMD simulation, we conducted experiments to detect THz-driven vibronic quantum beats $\Delta R_{osc}/R$, similar to Fig. 1b, by employing three different probe wavelengths: 750 nm, 755 nm, and 760 nm. The selection of these probe wavelengths was deliberate, aiming to match the threefold Rashba excitonic fine structure of MAPbI$_3$ sample, as shown in Figs. 2a-2b, with the degenerate dark exiton states ($\sim$755 nm, labeled as D) situated between two bright ones ($\sim$750 nm and 760 nm, labeled as B1 and B2) \cite{liu1}. 
Under these controlled experimental conditions, the vibronic quantum beat enables comprehensive investigation into the time-dependent buildup and decay of various phonon modes.
We then perform continuous wavelet transformation (CWT) of at probe wavelength 750 nm (Fig. 2c), 755 nm (Fig. 2e) and 760 nm (Fig. 2g) for the time delay $\Delta t_{pp}$ up to $\sim$ 33 ps at 4.2 K. The CWT approach (Supplemental Note 6) is a powerful time-frequency transformation that allows us to visualize how phonon modes evolve over time, revealing their development at different time intervals. To achieve high time resolution, the spectral resolution was compromised. The corresponding time-integrated power spectra obtained are shown in Figs. 2d, 2f, and 2h. 

The vibronic quantum beat spectra illustrated in Fig. 2 clearly show a mode- and symmetry-selective coupling of phonon modes to the three-fold excitonic fine structure splitting in MAPbI$_3$ and optical transition matrix elements associated with different exciton states \cite{liu1}. Specifically, this coupling occurs between two bright exciton states at approximately 750 nm and 760 nm, primarily interacting with multiple infrared (IR) modes $\sim$2-3 THz, and one dark exciton state at approximately 755 nm, predominantly engaging with Raman modes such as $\omega\mathrm{_6}=$0.8 THz. Plotting data together from three representative probe wavelengths for both bright (Figs. 2c and 2g) and dark states (Fig. 2e), allows us to visualize the timing of various vibronic modes' appearance and diminishment without the symmetry limitation of phonon modes, which are dependent on the excitonic transitions used for probing.

Next we highlight four crucial points in Fig. 2, which together provide a comprehensive chronological order of the lattice vibrations and coherent entropy transfer induced by THz excitation in MAPbI$_3$. 
First, the observed variations in the appearing and diminishing times for each phonon mode, as shown by the time-frequency map Figs. 2c, 2e, and 2g, are in line with the simulation results in Fig. 1c. During the time delay $\Delta t_{pp} < 3$ ps (Figs. 2c and 2g), it's noteworthy that phonon modes $\omega_1$ and $\omega_2$ (marked with red labels) appear nearly simultaneously right after the THz pumping. Those phonon modes are short-lived, diminishing quickly within $\sim$ 1 ps (Fig. 2c). 
Interestingly, the higher-frequency IR phonon modes, $\omega_1$ and $\omega_2$, do not precisely align with the THz pump spectrum centered around approximately 1 THz. While our broadband, single-cycle THz pulse provides some level of resonant excitation for these modes, they can also be driven in a non-resonant manner, such as through polaronic electron-phonon coupling, and manifest themselves first in time (Fig. 2c) \cite{liu1}.
Second, While the phonon mode $\omega_3$ emerges alongside the $\omega_1$ and $\omega_2$ modes due to resonant THz pump excitation, it displays a delayed enhancement, becoming more prominent as phonon modes $\omega_1$ and $\omega_2$ diminish in intensity. This behavior suggests a coherent transfer of energy between these phonon modes, with some of the energy from $\omega_1$ and $\omega_2$ being transferred to enhance the amplitude of $\omega_3$. 
Third, when $\Delta t_{pp}>$ 3 ps, with the diminishing of higher frequency phonon modes, the low energy phonon mode $\omega_4$ appears. Besides, a small portion of the energy was transfer to the $\omega_5$. The phonon mode $\omega_4$ centered at 0.3 THz lasts $\sim$6~ps. The microscopic origin of the 0.3 THz is from the molecular rotational entropy \cite{ref1, moving}. 
Fourth, as mostly clearly shown in Fig. 2e, at long time $\Delta t_{pp}$ $>$8~ps the time-frequency map of $\Delta R_{osc}/R$ at 755 nm shows a most pronounced and long-lasting phonon mode $\omega_6$ centered at 0.8~THz that dominates the quantum beat spectra. This octahedra cage twist (TO) mode $\omega_6$ at 0.8 THz is well established in the low-temperature Raman measurement \cite{Frost}. 
The observed coherent transfer of discrete phonon modes, $\omega_1$ to $\omega_6$, which entails both an increase and a decrease in phonon mode energy, underscores the critical role of nonlinear phononics, extreme anharmonicity, and lattice softness in shaping the bi-directional entropy transfer pathway in MAPbI$_3$.

\begin{figure*}[tbp]
	\includegraphics[scale=0.7]{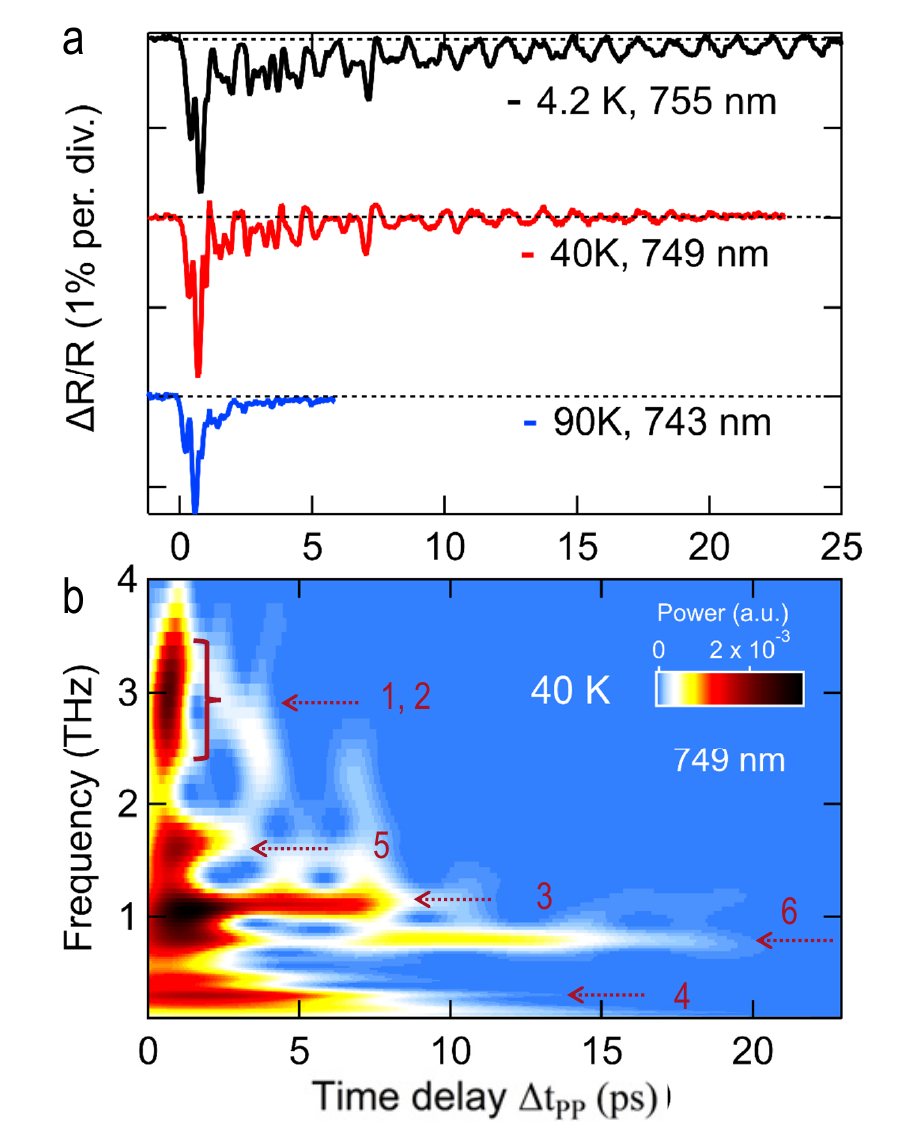}
	\caption{(a)Temperature-dependent $\Delta R/R$ dynamics at 4.2 K (black), 40 K (red), and 90 K(blue). The probe photon energy is adjusted to correspond with the temperature-dependent exciton transition as marked. (b) Continuous wavelet transform of THz pump-driven $\Delta R_{osc}/R$ dynamics at 40 K. The phonon modes  $\omega_1$--$\omega_6$ are labeled by the red arrows and numbers.}
	\label{Fig1}
\end{figure*}

The remarkable consistency observed between the experimental results of the phonon mode transfer sequence (Fig. 2) and the theoretical predictions (Fig. 1c) is noteworthy. While there are shorter time scales associated with the simulated phonon cascade process than experiment, these can be attributed to factors such as the precise parameters used in the simulation and the absence of energy dissipation in microcanonical ensemble simulations. Overall, this consistency reaffirms the validity of the essential dynamics of the coherent entropy transfer pathway in MAPbI$_3$ captured by both the theoretical framework and experiment. 
Additionally, the phonon shift behavior reveals an extreme nonlinear coupling spanning frequencies from approximately 3 THz down to 0.3 THz. This ratio exceeds those observed in other representative semiconductors, such as Ti$_2$O$_3$, by approximately 15 times (Supplemental Note 7). \cite{Ti2O3}. 
%

The coherent phonon entropy transfer can exhibit high sensitivity to temperature, primarily due to thermally-induced phonon damping and intensity reduction \cite{Bourges}. In Fig. 3a, we reveal such thermally-induced dephasing by plotting $\Delta R/R$ dynamics at three temperatures: 4.2 K (black), 40 K (red), and 90 K (blue).
When compared to 4.2 K, the dephasing time of phonon quantum oscillation significantly shortens at 40 K, diminishing within 20 ps. At 90 K, the long-lasting coherence vanishes.
To understand the impact of temperature on the time evolution of various phonon modes, we conduct the time-frequency analysis of the oscillatory residual $\Delta R_{osc}/R$ (Supplemental material) at 40 K and the result is shown in Fig. 3b. The decoherence time of phonon modes $\omega_1$ and $\omega_3$ remains unchanged, while the decoherence time of phonon mode $\omega_6$ at 0.8 THz is 10 ps shorter at 40 K compared to 4.2 K. This observation suggests the temperature sensitivity of the octahedra cage twist in the orthorhombic phase of MAPbI$_3$.
Despite the reduction in phonon lifetime, the phonon entropy transfer pathway at 40 K largely resembles that at 4.2 K. Specifically, phonon entropy continues to be transferred from high-frequency phonon modes $\omega_1$ to phonon modes $\omega_3$ and $\omega_4$ before dissipating through phonon mode $\omega_6$ at 0.8 THz.

We attribute the unique phonon entropy transfer process discovered to the gigantic structure anharmonicity of MAPbI$_3$ and the rich rotational dynamics of the methylammonium (MA) cation that interact to create the unique and intriguing coherent entropy transfer behavior. The uniqueness of hybrid organic-inorganic perovskites, when compared to tradition semiconductor materials such as Si or GaAs, lies in the large anharmonicity of inorganic perovskite lattice and the rich phonons of organic cations and their mutual coupling. The origins of the low-frequency phonon modes observed in Fig. 1b can be attributed to vibrational entropy stemming from the perovskite lattice vibrations and MA molecular liberation \cite{moving}. The combination of low stiffness and significant anharmonicity fosters electron-phonon collisions and phonon entropy transfer processes. These unique attributes not only serve as prerequisites for initiating lattice vibrations through photon pumping and polaronic coupling at frequencies around 1-3 THz ($\omega_1$, $\omega_2$ and $\omega_3$) but also play pivotal roles in facilitating coherent phonon transfer to modes below 1 THz thereafter. A diverse phonon band ranging from 0.3 to 0.7 THz is associated with the rotational motions of the MA organic cations \cite{Frost,ref6,moving}. The MA cation significantly participates in coherent and bi-directional entropy transfer. For instance, as evidenced in Figs. 2c, 2e and 2g, the low-frequency phonon at 0.3 THz ($\omega_4$) persists for approximately 6 ps before diminishing, transferring entropy to a higher-frequency perovskite cage mode at 0.8 THz ($\omega_6$).

\begin{figure*}[tbp]
	\includegraphics[scale=0.45]{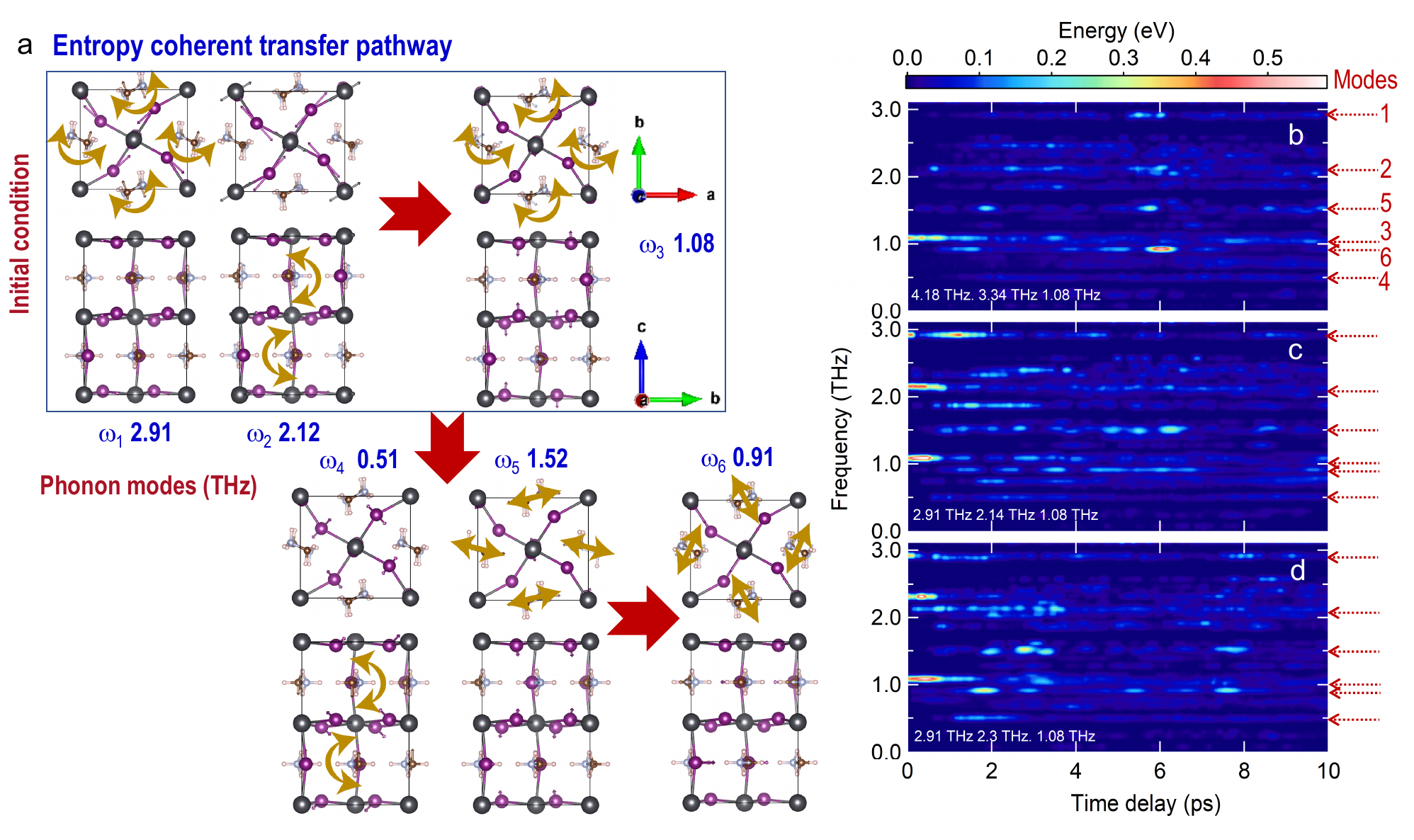}
	\caption{ (a) Schematic representation of the phonon entropy transfer pathway of MAPbI$_3$ identified experimentally in Fig. 2. The displacements of Pb, I, C, and N atoms are indicated by arrows, and the displacements of the H atoms are not shown for clarity. The directional movements of MA molecules, encompassing both rotational and shearing modes, are delineated by golden double-headed arrows. The time-evolution of the phonon mode-selective energy distribution function of the system throughout the 10 ps FPMD window with different initial phonon population at (b) 4.18, 3.34, 1.08 THz; (c),2.91, 2.14, 1.08 THz; (d) and 2.91, 2.3, 1.08 THz. The color scale indicates the energy of each phonon mode. The $\omega_1$--$\omega_6$ phonon modes involved are labeled by arrows and numbers 1-6.}
	
\end{figure*}

A unified picture emerges for the bi-directional entropy transfer pathway in MAPbI$_3$ by comparing the results of the FPMD simulation (Fig. 1c) and time-resolved vibronic quantum beat experiments (Fig.~2).  
In Fig. 4a, we provide a schematic representation of the phonon modes, emphasizing the directions of librations and translations of methylammonium molecules using golden double-headed arrows. These arrows correspond to the MA rotational mode and shearing mode, respectively, providing insights into the rich phononic entropy transfer originating from the inherent nonlinear phonon-phonon coupling effects of the material. 
Briefly, following THz excitation, phonon modes $\omega_1$ and $\omega_2$ were rapidly generated. These modes diminished within 1 ps as phonon mode $\omega_3$ was enhanced. These three modes constitute the initial conditions that govern entropy transfer over extended durations, as depicted inside the box of Figure 4a. The majority of entropy was transferred to the low-frequency phonon mode $\omega_4$ with only a small portion going to $\omega_5$. Coherence beats dissipated through the long-lasting phonon mode $\omega_6$, likely due to its primary coupling with dark excitons that appear $\sim$775nm as illustrated in Fig. 2b.

For a microscopic understanding of the phonon dynamics, it's important to note that certain phonon modes, like mode $\omega_4$ at 0.51 THz and mode $\omega_3$ at 1.08 THz, already exhibit noticeable anharmonicity in their single-mode potential energy surfaces (PES), as illustrated in the supplemental note 5 (Fig. S5).
To provide a quantitative analysis, we further assess the phonon-phonon coupling coefficients by mapping out the PES along two or three phonon mode coordinates and fitting them with higher-order cubic and quartic terms, as outlined in Eqs. S3 and S4 (supplemental note 5). The description of multidimensional PESs transcends harmonic terms and implies significant nonlinear inter-phonon couplings, as demonstrated in the Supplementary materials (Fig. S6).
Specific fitting parameters can be found in Tables S1 and S2 (supplemental note 5), highlighting appreciable inter-mode couplings among all relevant phonons. This analysis reveals the complex and nonlinear nature of the interactions between phonon modes within the MAPbI$_3$ material. 

The extreme nonlinear phononics observed are expected to be sensitive to the initial conditions.  
The first-principle molecular quantum dynamics simulation serves a dual purpose: it unravels the intricate pathways of coherent coupling and also identifies the crucial roles played by the initial phonon modes induced by THz excitation.
To further investigate this sensitivity to initial conditions, we compare the FPMD simulations using three different sets of phonon modes, distinct from those obtained experimentally, as initial conditions. This comparative analysis allows us to explore the impact of different starting configurations on the observed coherent phonon dynamics and entropy transfer pathway.
We maintain the phonon mode $\omega_3$ at 1.08 THz, originating from direct THz excitation, while varying the phonon modes $\omega_1$ and $\omega_2$, permitted by crystal symmetry, to represent the initial high density of phonons.
The results are shown in Figs. 4b to 4d. 
On the one hand, in our initial simulations, we chose phonon mode $\omega_1$ at 4.18 THz and mode $\omega_2$ at 3.34 THz, which were not observed in the experimentally measured quantum beat spectra in Fig. 1b. While all the relevant phonon modes can still be excited with these initial conditions, the mode evolution sequence in Fig. 4b differs significantly from the experimental observations. 
Notably, in this scenario, phonon mode $\omega_4$ emerges subsequent to the excitation of phonon mode $\omega_6$, displaying a significantly lower intensity compared to the experimental findings illustrated in Figure 2. This observation suggests that the progression of phonon modes is not stochastic but rather intricately influenced by the initial phonon distributions.
On the other hand, when we maintain phonon mode $\omega_1$ at 2.91 THz and select mode $\omega_2$ at 2.14 THz (Fig. 4c) and 2.3 THz (Fig. 4d), the results still differ from those in Fig. 1c. In these cases, the final transferred energy intensity at $\omega_6$ is notably weaker, and the dissipation occurs more rapidly compared to the behavior observed in Fig. 1c. This corroborates the sensitivity of the phonon dynamics and entropy transfer pathway to the specific initial conditions for the simulation.

It is worth further delineating entropy vs energy transfer for the observed bi-directional phonon mode evolution. Comparing the initial conditions chosen for Fig. 1c (2.91 THz, 2.12 THz, and 1.08 THz) and Fig. 4c (2.91 THz, 2.14 THz, 1.08 THz), the phonon energy scales are essetially the same. However, the major difference originates from the fact that the mode at 2.14 THz and that at 2.12 THz, although indeed are very close in frequency, but are very different eigen-modes (Supplemental Note 8). In other words, the associated atomic vibration patterns are entirely different (the eigenvectors are orthogonal). This precisely illustrates that entropy transfer stemming from atomic distribution isn't solely dictated by mode energy. Hence, the frequency alone doesn't provide a comprehensive picture of the atomic motion of phonon modes and their anharmonicity. This highlights the significance of our time-frequency quantum beat spectroscopy in discerning entropy transfer processes.

It is vitally important to comprehend and regulate phonon entropy transfer, coherence, and lattice anharmonicity in perovskite-based photovoltaic and optoelectronic applications. First, the stochastic cation reorientations cause the dipolar scattering to charge carriers. They are highly detrimental to the transport, migration and recombination of charge carriers, factors that govern carrier mobility and diffusion length in photovoltaic devices.
The insights derived from the dynamic entropy transfer pathway, as revealed by our THz vibronic quantum beat spectroscopy measurements, offer a valuable framework for measuring and understanding strategies to minimize the entropy-related losses. This control is crucial for enhancing the overall efficiency of the photovoltaic device.
Second, key parameters of light-emitting devices, such as spectral width and internal quantum efficiency, are significantly influenced by carrier-phonon interaction and phonon-phonon nonlinear dynamics. The pronounced anharmonicity and dynamic fluctuations unveiled in our perovskite study underscore a distinct contrast from conventional semiconductors. Therefore, our results provide a new avenue for materials design and discovery in advancing perovskite-based semiconductor optoelectronics.
Third, our experimental and theoretical findings suggest the possibility of inducing mode-specific, long-lasting lattice coherence. This discovery introduces a novel method for directing charge transfer and collection by mitigating thermal phonon scattering and enhancing coherence in hybrid perovskites.

In summary, we discover a coherent and bi-directional transfer of entropy in the orthorhombic phase of MAPbI$_3$ by combining THz-driven quantum beat spectroscopy and first-principles molecular quantum dynamics simulation. 
This approach proves to be a powerful and versatile tool for measuring dynamic phonon entropy transfer, coherence, and dissipation in metal halide perovskites. 
Intriguing theoretical future avenues encompass explicit exploration of light-matter interactions to ascertain initial phonon populations and the integration of dissipation into dynamic simulations.
The discovery of coherent entropy transfer holds the promise of offering a systematic material design strategy to enhance coherence, dynamics, and lattice anharmonicity, thus driving forward the development of high-performance perovskite-based photovoltaic and optoelectronic applications.

\bibliographystyle{Science}

\section*{Acknowledgments}
This work was supported by the Ames National Laboratory, the US Department of Energy, Office of Science, Basic Energy Sciences, Materials Science and Engineering Division under contract No. DE-AC02-07CH11358 (Scientific lead, THz spectroscopy and model building). Sample development in The University of Toledo were supported by the U.S. Air Force Research Laboratory, Space Vehicles Directorate, under Contract No. FA9453-19-C-1002. First-principles molecular dynamics simulations were performed at Hefei Advanced Computing Center and Supercomputing center at USTC.



\section*{Correspondence} Correspondence and requests for materials should be addressed to J.W. (jgwang@ameslab.gov;
jgwang@iastate.edu).






\clearpage


\end{document}